\documentclass[12pt]{article}
\usepackage{amsmath}
\usepackage{amssymb}
\usepackage{amsfonts}
\usepackage{amscd}
\usepackage{delarray}
\usepackage{graphicx}

\usepackage{amsmath,amsmath,amsfonts,amssymb,color,bbm}

\usepackage{graphicx}

\def\Symp#1,#2,#3,#4.{\left[\left(\begin{array}{c}#1\\#2\end{array}\right),\left(\begin{array}{c}#3\\#4\end{array}\right)\right]}
\def\Vec#1,#2.{\left(\!\begin{array}{c}#1\\#2\end{array}\!\right)}
\def\vec#1,#2.{{#1\choose{#2}}}
\def\ket#1.{|#1\rangle}

\def\bra#1.{\langle#1|}
\def\braket#1,#2.{\langle#1|#2\rangle}
\newcommand{\beq}{\begin{equation}}
\newcommand{\eeq}{\end{equation}}
\newcommand{\beqa}{\begin{eqnarray}}
\newcommand{\eeqa}{\end{eqnarray}}

\usepackage{soul}
\definecolor{purplerep}{rgb}{1,0.1,1}

\definecolor{green}{rgb}{0.2,0.6,0.8}

\begin{document}

\title{Testing de Broglie's double solution in the mesoscopic regime.}
\date{}
\author{}
\maketitle

\centerline{Thomas Durt\footnote{Aix Marseille Universit\'e, CNRS, Centrale Marseille, Institut Fresnel, UMR 7249, 13013 Marseille, France; e-mail: thomas.durt@centrale-marseille.fr}}

\begin{abstract}

We present here solutions of a non-linear Schr\"odinger equation in presence of an arbitrary linear external potential. The non-linearity expresses a self-focusing interaction. These solutions are the product of the pilot wave with peaked solitons the velocity of which obeys the guidance equation derived by Louis de Broglie in 1926. The degree of validity of our approximations increases when the size of the soliton decreases and becomes negligible compared to the typical size over which the pilot wave varies. We discuss the possibility to reveal their existence by implementing a humpty-dumpty Stern-Gerlach interferometer in the mesoscopic regime.
\end{abstract}
Keywords: de Broglie-Bohm dynamics; double solution program, self-interaction, gravitation.

\section{Introduction.}
In continuation of a previous paper \cite{debroglieDurt} we consider here modifications of the linear Schr\"odinger equation due to the hypothetical existence of a non-linear potential expressing the self-interaction of the quantum particle. 
The modified evolution equation then reads\begin{equation}
{i}\hbar\frac{\partial\Psi(t,{{\bf x}})}{\partial t}=-\hbar^2\frac{\Delta\Psi(t,{{\bf x}})}{2m}
+V^L(t,{{\bf x}})\Psi(t,{{\bf x}})+V^{NL}(\Psi)\Psi(t,{{\bf x}}),\label{nonfreeNL}
\end{equation}
where $V^L$ represents an arbitrary linear potential, of the type commonly considered when solving the linear Schr\"odinger equation (for instance an electro-magnetic potential) while $V^{NL}$ represents a non-linear self-focusing potential which supposedly concentrates the wave function of the particle over a tiny region of space, in accordance with de Broglie's double solution program.
 We also impose the factorization ansatz:
\begin{equation}\Psi(t,{{\bf x}})=\Psi_L(t,{{\bf x}})\cdot \phi_{NL}(t,{{\bf x}}), \label{ansatz}\end{equation}
 where $\Psi_L$, the so-called ``pilot'' wave, is a solution of the linear Schr\"odinger equation:  \begin{equation}i\hbar{\partial\over \partial t}\Psi_L({\bf x},t)=({-\hbar^2\over 2m}\nabla^2 +V^L(x,y,z,t))\Psi_L({\bf x},t),\label{S1V}\end{equation} while $\phi_{NL}(t,{{\bf x}})$ is supposed to be localized over a very small region of space.
 Our aim is to represent the particle by $ \phi_{NL}(t,{{\bf x}})$, a soliton guided by the pilot wave according to de Broglie's guidance equation:
 \begin{equation}{\bf v}_{dB-B}={\hbar \over m}{Im.(\Psi_L({\bf x},t)^*{\bf \nabla}\Psi_L({\bf x},t))\over |\Psi_L({\bf x},t)|^2}.\label{guidance}\end{equation}
 de Broglie presented the guidance equation (\ref{guidance}) in the Solvay conference of 1927 \cite{1927}, together with an interpretation which constitutes the so-called de Broglie-Bohm interpretation  \cite{bohm521,bohm522,Holland}, sometimes called the pilot-wave interpretation, a simplified version of de Broglie's double solution program. According to the pilot-wave interpretation particles are localized in tiny regions of space at any time and follow continuous trajectories in accordance with the guidance equation (\ref{guidance}) which expresses how the linear (pilot) wave guides the particles. According to de Broglie's double solution program \cite{debrogliebook,debroglieend,fargueannales}, a supplementary constraint should be imposed, which is, roughly summarized:

{\it Particles are waves, and are identified with solutions of a modified non-linear Schr\"odinger equation, here denoted $\Psi(t,{{\bf x}})$ (see equation (\ref{ansatz})).  These solutions are solitons, that is to say waves for which spread gets compensated by the (self-focusing) non-linearity. }
  
  The factorization ansatz (\ref{ansatz}) is a novelty of our approach \cite{debroglieDurt}. It results from the recognition that, due to the fundamental non-linearity of the wave dynamics, a linear partition of the type $\Psi(t,{{\bf x}})=\Psi_L(t,{{\bf x}})+ \phi_{NL}(t,{{\bf x}})$ as was considered by de Broglie \cite{debrogliebook} is irrelevant. From this point of view the factorization ansatz incorporates non-linearity from the beginning.
  
  As we shall show in section 2, a well-chosen non-linearity makes it possible to derive solutions of equations (\ref{nonfreeNL},\ref{ansatz},\ref{S1V}) whose trajectories obey the guidance equation (\ref{guidance}). These are in general approached solutions, and the degree of validity of our approximations is shown to increase when the size of the soliton decreases. When the external potential is harmonic (quadratic) and when the linear wave represents a coherent state, our approach delivers an exact solution however, as we shall show in the same section. In section 3, we show how to generalize our approach to the many particles case. In section 4, we incorporate the gravitational interaction into our model and propose an experiment aimed at revealing the existence of solitonic solutions similar to those described here, thanks to an interferometric humpty-dumpty Stern-Gerlach experiment. We conclude in the last section.

\section{Single particle case.}
\subsection{Preliminary results.}

Combining equations (\ref{nonfreeNL},\ref{ansatz},\ref{S1V}), expressing $ \Psi_L(t,{{\bf x}})$ in fuction of its modulus and its phase as $R_L(t,{{\bf x}})e^{i\varphi_L(t,{{\bf x}})}$, and also making use of the identity ${\bf \bigtriangledown}   \Psi_L(t,{{\bf x}})=({\bf \bigtriangledown}   R_L(t,{{\bf x}}))e^{i\varphi_L(t,{{\bf x}})}+\Psi_L(t,{{\bf x}})i{\bf \bigtriangledown}\varphi_L(t,{{\bf x}})$, it is straightforward to show that $ \phi_{NL}$ obeys the non-linear equation
  \beqa &&{i}\hbar\cdot  \frac{\partial \phi_{NL}(t,{{\bf x}})}{\partial t}=\nonumber
-\frac{\hbar^2}{2m}\cdot \Delta\phi_{NL}(t,{{\bf x}}) \\& -&\frac{\hbar^2}{m}\cdot (i{\bf \bigtriangledown}  \varphi_L(t,{{\bf x}}) \cdot {\bf \bigtriangledown} \phi_{NL}(t,{{\bf x}})+\frac{{\bf \bigtriangledown}  R_L(t,{{\bf x}})}{R_L(t,{{\bf x}})} \cdot   {\bf \bigtriangledown}  \phi_{NL}(t,{{\bf x}}))\nonumber\\ &+&V^{NL}(\Psi)\phi_{NL}(t,{{\bf x}})\label{S2}, \eeqa

By doing so we replace thus equation (\ref{nonfreeNL}) by a system of three equations (\ref{ansatz},\ref{S1V},\ref{S2}). This replacement is one to one and can be done without loss of generality whenever ${\bf x}$ is not a node of the pilot-wave $\Psi_L(t,{{\bf x}})$ which is the case ``nearly everywhere''.

   In order to solve the system of equations (\ref{S1V},\ref{S2}), it is worth noting that while the L$_2$ norm of the linear wave $ \Psi_L$ is preserved throughout time, because (\ref{S1V}) is unitary, this is no longer true in the case of the non-linear wave $\phi_{NL}$, because the terms mixing $\Psi_{L}$ and $\phi_{NL}$ in  (\ref{S2}) are not hermitian. The change of norm of $\phi_{NL}$ can be shown  \cite{debroglieDurt}, to obey
   \beqa{d <\phi_{NL}|\phi_{NL}>\over dt}\approx \frac{\hbar}{m}{ \Delta}  \varphi_L(t,{{\bf x_0}}) \cdot <\phi_{NL}    |\phi_{NL}> -2\frac{{\bf \bigtriangledown}  R_L(t,{{\bf x_0}})}{R_L(t,{{\bf x_0}})}\cdot 
\int d^3{\bf x} (\phi_{NL}(t,{{\bf x}}))^*\frac{\hbar\bf \bigtriangledown} {mi } \cdot     \phi_{NL}(t,{{\bf x}}),\label{normchangebis}\eeqa
where we introduced the barycentre ${\bf x_0}$ of the soliton: ${\bf x_0}\equiv {  <\phi_{NL}| {\bf x}|\phi_{NL}>\over <\phi_{NL}|\phi_{NL}>}.$
Remark that the bra-ket notation introduced here should not necessarily be interpreted as a quantum statistical average in the usual sense; it rather indicates an average quantity in regard of the weight (density of stuff) $|\phi_{NL}(t,{\bf x} )|^2$. 
   
     Still in  reference \cite{debroglieDurt}, we established the following results (i,ii): 
   
   i) If we define the velocity ${\bf v}_{drift}$ of the barycentre ${\bf x_0}$ as follows: \beqa {\bf v}_{drift}\equiv{d({  <\phi_{NL}| {\bf x}|\phi_{NL}>\over <\phi_{NL}|\phi_{NL}>} ) \over dt       }\label{defdrift}\eeqa
   Then,
  \beqa {\bf v}_{drift}&=&{\hbar \over m}{{\bf \bigtriangledown}}\varphi_L({{\bf x_0}}(t),t)+{<\phi_{NL}| {\hbar\over i m}{\bf \bigtriangledown}|\phi_{NL}>\over <\phi_{NL}|\phi_{NL}>}\label{drift} \\&=&{\bf v}_{dB-B}+{\bf v}_{int.}.\nonumber\eeqa which contains the well-known Madelung-de Broglie-Bohm contribution (${\bf v}_{dB-B}={\hbar \over m}{{\bf \bigtriangledown}}\varphi_L({{\bf x_0}}(t),t)$) plus a new contribution due to the internal structure of the soliton (${\bf v}_{int.}={<\phi_{NL}| {\hbar\over i m}{\bf \bigtriangledown}|\phi_{NL}>\over <\phi_{NL}|\phi_{NL}>}$).

ii)   In the limit where the soliton is peaked enough around its barycentre:  \beqa{<\phi_{NL}|\phi_{NL}>(t)\over <\phi_{NL}|\phi_{NL}>(t=0)}={R^2_L({\bf x_0},t=0)\over R^2_L({\bf x_0},t)},\label{defnorm}\eeqa where ${\bf x_0}(t)$, the barycentre of $\phi_{NL}$, moves according to the generalized dB-B guidance equation (\ref{drift}).
   
The first result (i) strongly suggests the possible existence of a purely real solitonic solution of equation ($\ref{S2}$) such that ${\bf v}_{int.}=0$ in which case the guidance equation of de Broglie (\ref{guidance}) is satisfied: ${\bf v}_{drift}={\hbar \over m}{{\bf \bigtriangledown}}\varphi_L({{\bf x_0}}(t),t)$. Having in mind that all aforementioned results were derived in the limit where the width of the peaked soliton is quite smaller than the typical scales of variation of $R_L(t,{{\bf x}}))$ and $\varphi_L(t,{{\bf x}})$ over space, the second result (ii) implies that the full wave function $\Psi$ solution of  (\ref{S2}) has the form 
  \begin{equation}\Psi(x,y,z,t) \approx \phi'_{NL}({\bf x},t)e^{i\varphi_L({\bf x},t)},\end{equation}  where $\phi'_{NL}({\bf x},t)$ is centered in $  {\bf x}_0(t=0)+\int_0^t dt {\bf v}_{drift}$ and is of constant L$_2$ norm.  
  
   
\subsection{A formal realization of de Broglie's double solution program.}
 
 In order to represent the particle by $ \phi_{NL}(t,{{\bf x}})$, a soliton guided by the pilot wave according to de Broglie's guidance equation (\ref{guidance}), let us make the following choice for $V^{NL}$:
 
 \begin{equation}V^{NL}(\Psi)={\hbar^2\over 2m}{\Delta|\Psi(t,{{\bf x}})|\over |\Psi(t,{{\bf x}})|}-{\hbar^2\over 2m}{\Delta|\Psi_L(t,{{\bf x}})|\over |\Psi_L(t,{{\bf x}})|}\end{equation}

As has been shown by Bohm \cite{bohm521,bohm522,Holland}, making use of the linear Schr\"odinger equation (\ref{S1V}), when the guidance condition (\ref{guidance}) is fulfilled, the acceleration of the barycentre of the quantum particle obeys Newton's equation 

$m{\bf a}({\bf x_0},t)=m{d {\bf v}_{dB-B}({\bf x_0},t)\over dt}=-{\bf \nabla} V^L(({\bf x_0},t))-{\bf \nabla} V^Q_L(({\bf x_0},t))$ where $V^Q_L(({\bf x_0},t))$ represents the (non-linear) quantum potential:

 \begin{equation}V^Q_{L}(\Psi_L)=-{\hbar^2\over 2m}{\Delta|\Psi_L(t,{{\bf x}})|\over |\Psi_L(t,{{\bf x}})|}\end{equation}
 
 Here, we introduce a non-linear potential which is the difference between $V^Q_{L}(\Psi_L)$ the ``usual'' quantum potential, associated to the pilot wave, with a quantum potential associated to the full wave function  $V^Q(\Psi)=-{\hbar^2\over 2m}{\Delta|\Psi(t,{{\bf x}})|\over |\Psi(t,{{\bf x}})|}$.
 
 Making use of the factorisation ansatz (\ref{ansatz}) we can rewrite $V^{NL}(\Psi)$ as follows:
 
 \begin{equation}V^{NL}(\Psi)=V_L^Q(\Psi_L)-V^Q(\Psi)={\hbar^2\over m}{\nabla |\Psi_L(t,{{\bf x}})|\over |\Psi_L(t,{{\bf x}})|}\cdot{\nabla|\phi_{NL}(t,{{\bf x}})|\over |\phi_{NL}(t,{{\bf x}})|}+{\hbar^2\over 2m}{\Delta|\phi_{NL}(t,{{\bf x}})|\over |\phi_{NL}(t,{{\bf x}})|}. \label{S3}\end{equation}
 
 The potential ${\hbar^2\over 2m}{\Delta|\phi_{NL}(t,{{\bf x}})|\over |\phi_{NL}(t,{{\bf x}})|}$ plays the role of a self-focusing potential but is non-accelerating as we shall show soon. The role of the potential ${\hbar^2\over m}{\nabla| \Psi_L(t,{{\bf x}})|\over |\Psi_L(t,{{\bf x}})|}\cdot{\nabla|\phi_{NL}(t,{{\bf x}})|\over |\phi_{NL}(t,{{\bf x}})|}$ is more subtle: in combination with  the non-hermitian terms mixing $\Psi_{L}$ and $\phi_{NL}$ in equation (\ref{S2}), it contributes to the de Broglie-Bohm self-acceleration. Indeed, combining equations (\ref{S2}) and (\ref{S3}), and making use of the fact that $|\Psi_L(t,{{\bf x}})|=R_L(t,{{\bf x}})$, we get 
 
   \beqa &&{i}\hbar\cdot  \frac{\partial \phi_{NL}(t,{{\bf x}})}{\partial t}=\label{S4}
-\frac{\hbar^2}{2m}\cdot (\Delta\phi_{NL}(t,{{\bf x}})- \Delta |\phi_{NL}(t,{{\bf x}})|\cdot {\phi_{NL}(t,{{\bf x}})\over |\phi_{NL}(t,{{\bf x}})|})\\& -&\frac{\hbar^2}{m}\cdot (i{\bf \bigtriangledown}  \varphi_L(t,{{\bf x}}) \cdot {\bf \bigtriangledown} \phi_{NL}(t,{{\bf x}})+\frac{{\bf \bigtriangledown}  R_L(t,{{\bf x}})}{R_L(t,{{\bf x}})} \cdot   ({\bf \bigtriangledown}  \phi_{NL}(t,{{\bf x}})-{{\bf \bigtriangledown}  |\phi_{NL}(t,{{\bf x}})|\over|  \phi_{NL}(t,{{\bf x}})|}\cdot \phi_{NL}(t,{{\bf x}})).\nonumber \eeqa

This equation grandly simplifies whenever $\phi_{NL}(t,{{\bf x}})$ is a real positive function, in which case equation (\ref{S4}) reads

 \beqa &&{i}\hbar\cdot  \frac{\partial \phi_{NL}(t,{{\bf x}})}{\partial t}=-\frac{\hbar^2}{m}\cdot i{\bf \bigtriangledown}  \varphi_L(t,{{\bf x}}) \cdot {\bf \bigtriangledown} \phi_{NL}(t,{{\bf x}})\label{S5}. \eeqa

If the size of the soliton is quite smaller than the typical size of variation of ${\bf \bigtriangledown}  \varphi_L(t,{{\bf x}})$, that is to say, if it is quite smaller than ${||{\bf \bigtriangledown} \varphi_L(t,{{\bf x_0}})||\over |\Delta \varphi_L(t,{{\bf x_0}})|}$, we may replace, in good approximation, ${\bf \bigtriangledown}  \varphi_L(t,{{\bf x}})$ by ${\bf \bigtriangledown}  \varphi_L(t,{{\bf x_0}}) $. Then, it is easy to solve equation (\ref{S5}) for which we find a solitary wave solution of the type

\begin{equation} \phi_{NL}(t,{{\bf x}})= \phi_{NL}({\bf x}-{{\bf x_0}}(t)),\label{guidancebis}\end{equation}with ${{\bf x_0}}(t)={{\bf x_0}}(t_0)+\int_{t_0}^tdt {\bf v}_{dB-B}({{\bf x_0}}(t)).$ Remarkably, this solution is valid independently of the initial shape of $\phi_{NL}$ at time $t_0$, provided it is a real positive function of the position. It moves without deformation at all times and remains thus a real positive function, moving at the velocity ${\bf v}_{dB-B}$, in accordance with equation (\ref{drift}) because when $\phi_{NL}$ is real, ${\bf v}_{int.}={<\phi_{NL}| {\hbar\over i m}{\bf \bigtriangledown}|\phi_{NL}>\over <\phi_{NL}|\phi_{NL}>}=0$.

\subsection{Guidance equation associated to the coherent state of an harmonic oscillator: an exact solution.}

Let us consider in particular that the linear potential is harmonic and that the pilot wave is a coherent state:
\begin{equation} \Psi_L(t,{{\bf x}})=({m\omega\over \pi \hbar})^{1/4} \exp\left[-({m\omega\over \hbar})({\bf x}-{\bf \tilde x_0}cos(\omega t))^2+i\sqrt{{2m\omega\over \hbar}}{\bf \tilde x_0}sin(\omega t){\bf x}+i\theta(t)\right]\end{equation}

Then, equation (\ref{S4}) is gaussian which means that if we impose that $\phi_{NL}$  is gaussian at time $t=t_0$ it will remain gaussian for all times which grandly simplifies the computations.

 Accordingly, let us try the following expression for $\phi_{NL}$: $\phi_{NL}(t,{\bf x})=$

$\exp\left[-A_x(t) \frac{x^2}{2}+B_x(t)x+C_x(t)\right]\exp\left[-A_y(t) \frac{y^2}{2}+B_y(t)y+C_y(t)\right]\exp\left[-A_z(t) \frac{z^2}{2}+B_z(t)z+C_z(t)\right]$ where $A(t)$, $B(t)$ and $C(t)$ are complex functions of time.


Then ${\hbar^2\over 2m}{\Delta|\phi_{NL}(t,{{\bf x}})|\over |\phi_{NL}(t,{{\bf x}})|}= {\hbar^2\over 2m}((ReA_x(t))^2(x- x_0)^2+(ReA_y(t))^2(y-y_0)^2+(ReA_z(t))^2(z-z_0)^2)$,
up to an irrelevant additive constant, so that equation (\ref{S2}) becomes
 \beqa &&{i}\hbar\cdot  \frac{\partial \phi_{NL}(t,{{\bf x}})}{\partial t}=\label{S6gaussf}
-\frac{\hbar^2}{2m}\cdot \Delta\phi_{NL}(t,{{\bf x}}) -\frac{\hbar^2}{m}\cdot i{\bf \bigtriangledown}  \varphi_L(t,{{\bf x}}) \cdot {\bf \bigtriangledown} \phi_{NL}(t,{{\bf x}})\\ \nonumber&+&{\hbar^2\over 2m}((ReA_x(t))^2(x- x_0)^2+(ReA_y(t))^2(y-y_0)^2+(ReA_z(t))^2(z-z_0)^2))\phi_{NL}(t,{{\bf x}}),\eeqa  which puts into evidence the self-focusing nature of the purely solitonic part of the non-linear potential (${\hbar^2\over 2m}{\Delta|\phi_{NL}(t,{{\bf x}})|\over |\phi_{NL}(t,{{\bf x}})|}$). This potential is not self-accelerating as can be shown from a straightforward application of Ehrenfest's theorem (see e.g.  Ref.\cite{Biali} for a similar result derived in the framework of the logarithmic non-linear Schr\"odinger equation: due to the parity of the gaussian function under reflexions around its barycentre, the global contribution of the non-linear potential (here of ${\hbar^2\over 2m}{\Delta|\phi_{NL}(t,{{\bf x}})|\over |\phi_{NL}(t,{{\bf x}})|}$) to the self-acceleration vanishes). 

The equation (\ref{S6gaussf}) is also separable in cartesian coordinates  because the pilot wave is separable. Henceforth, to simplify the treatment, we shall from now on consider the evolution of the $x$ component only and again force  a gaussian solution of the type
\begin{equation}
\phi_{NL}(t,x)=\exp\left[-A_x(t) \frac{x^2}{2}+B_x(t)x+C_x(t)\right]\label{gaussf}
\end{equation}

The restriction of the self-interaction potential along $X$, ${\hbar^2(Re(A_x))^2\over 2m}(x-x_0)^2$, is peaked around the barycentre $x_0$ of the soliton and is equal to its Taylor development to the second order in $x$, of the form $V_{0}(t)+V_{1}(t)\,x+V_{2}(t)\,x^2$.

When the pilot wave is a coherent state, $\bigtriangledown_x \varphi_L(t,{{\bf x}})$ is exactly equal to $\bigtriangledown_x \varphi_L(t,{{\bf x_0}})$ because the phase of the pilot wave linearly depends on the position.

 We get thus after straightforward computations the following system of equations:\\ \\
\begin{equation}\large
\left\{ \begin{array}{rcl}
i\,\frac{d A_x(t)}{dt}&=&\frac{\hbar}{m}\,A_x(t)^2-2\,\frac{V_{2}(t)}{\hbar}=\frac{\hbar}{m}\,A_x(t)^2-
2\,{\hbar (Re(A_x))^2\over 2m}\\ \\
i\,\frac{d B_x(t)}{dt}&=&\frac{\hbar}{m}\,A_x(t)\,B_x(t)+\frac{V_{1}(t)}{\hbar}+i{\hbar\over m}A_x(t)\bigtriangledown_x \varphi_L(t,{{\bf x_0}})\\ \\
i\,\frac{d C_x(t)}{dt}&=&\frac{\hbar}{2\,m}\,\left[A_x(t)-B_x(t)^2\right]+\,\frac{V_{0}(t)}{\hbar}-i{\hbar\over m}B_x(t)\bigtriangledown_x \varphi_L(t,{{\bf x_0}}).
\end{array}\right.\label{gausg}
\end{equation}

 It admits solitonic solutions (similar to coherent states) for which $A_x=A_0$ (the same condition holds also for other cartesian axes of reference: $A_y=A_z=A_0$). Then the evolution of $B_x(t)$ considerably simplifies and we get
 
 \begin{equation}i\,\frac{d B_x(t)}{dt}=\frac{\hbar}{m}\,A_0\,B_x(t)+\frac{V_{1}(t)}{\hbar}+i{\hbar\over m}A_0\bigtriangledown_x \varphi_L(t,{{\bf x_0}}).\label{bof}\end{equation}

Now, $x_0=\frac{ \mathcal{R}\text{e}B_x}{ \mathcal{R}\text{e}A_x}=\frac{ \mathcal{R}\text{e}B_x}{ A_0}$
and $V_1(t)=-{\hbar^2\over m}A_0{ \mathcal{R}\text{e}B_x}$. Making use of equation (\ref{bof}) we also have $-\frac{d { \mathcal{I}\text{m}B_x}}{dt}=\frac{\hbar}{m}\,A_0\,\mathcal{R}\text{e}B_x(t)+\frac{V_{1}(t)}{\hbar}=0$ so that at all times the soliton remains real ($\mathcal{I}\text{m}B_x(t)=0$), provided the initial condition is a purely real gaussian function (all this up to an irrelevant global phase).

We also get $\frac{d { \mathcal{R}\text{e}B_x}}{dt}=\frac{\hbar}{m}\,A_0\,\bigtriangledown_x \varphi_L(t,{{\bf x_0}})$ but $\mathcal{R}\text{e}B_x= A_0\cdot x_0$ so that $\frac{d {x_0}}{dt}=\frac{\hbar}{m}\bigtriangledown_x \varphi_L(t,{{\bf x_0}})$.

Finally, making use of the previous results, we get $\frac{d \mathcal{R}\text{e} C_x(t)}{dt}=-{\hbar\over m}B_x(t)\bigtriangledown_x \varphi_L(t,{{\bf x_0}})=-A_0\cdot x_0\cdot \frac{d {x_0}}{dt}=A_0\frac{d {x^2_0}}{2dt}$ so that  $\mathcal{R}\text{e} C_x(t)$=$-A_0\frac{ {x^2_0}}{2}$ and $\mathcal{R}\text{e}(-A_x(t) \frac{x^2}{2}+B_x(t)x+C_x(t))=-A_0(x-x_0)^2$.

For solitons prepared at time $t=0$ in the state $\Psi(t=0,{{\bf x}})=\Psi_L(t=0,{{\bf x}})\cdot N\cdot e^{-{A_0\over 2}({\bf x}-{\bf x_0})^2}$, we predict thus that at time $t$, the state evolves to $\Psi(t,{{\bf x}})=\Psi_L(t,{{\bf x}})\cdot N\cdot e^{-{A_0\over 2}({\bf x}-{\bf x_0(t)})^2}$ with $\frac{d {\bf x_0}}{dt}=\frac{\hbar}{m}{\bf \bigtriangledown }\varphi_L(t,{{\bf x_0}})$. The guidance condition (\ref{guidance}) is thus satisfied. Actually all solitons follow trajectories of the type ${\bf x_0}(t)={\bf x_0}(t=0)-{\bf \tilde x_0}+{\bf \tilde x_0}cos(\omega t)$. They remain thus equidistant at all times with the peak of the pilot wave\footnote{Note that, to simplify the mathematical treatment we assumed here that the three cartesian components of the average position of the pilot wave oscillate in phase. Now, as the equation (\ref{S6gaussf}) is separable in cartesian coordinates, our results are still valid if we relax this assumption.}. This is due to the fact that the envelope of the pilot wave (a coherent state) moves without deformation. de Broglie-Bohm trajectories thus conspire in order to transport the initial position density as a whole. Remark that if initially the shape of the soliton is not gaussian it will also move without deformation and remain equidistant at all times from the peak of the pilot wave, because the solution (\ref{guidancebis}) is an exact solution when the pilot wave is a coherent state. 

 Another situation for which the solution (\ref{guidancebis}) is exact occurs when the linear potential vanishes; then plane pilot waves solutions of the linear Schr\"odinger equation (\ref{S1V})  are associated to a uniform movement of the soliton, which is a manifestation in this case of Galilei invariance \cite{debroglieDurt,CDWannales}. Other non-linear modifications of the Schr\"odinger equation such as e.g. the logarithmic non-linear Schr\"odinger equation \cite{Biali}, the NLS equation \cite{Fargue} or the Schr\"odinger-Newton equation \cite{CDW,lopez} also possess exact solutions in the form of the product of a plane wave modulation with a soliton moving at constant speed; such potentials can be shown however to be deprived of self-acceleration which prohibits their use for attempting to realize de Broglie's double solution program.

\section{The many particles case: guidance equation in configuration space.}
When more than one particle is present, a formal generalization of the previous results holds as we show now. To do so, we consider a wave function of the type

\begin{equation}\Psi(t,{{\bf x^1}},{\bf x^2},...{\bf x^i}...{\bf x^N})=\Psi_L(t,{{\bf x^1}},{\bf x^2},...{\bf x^i}...{\bf x^N})\cdot \phi_{NL}(t,{{\bf x^1}})\cdot\phi_{NL}(t,{{\bf x^2}})...\phi_{NL}(t,{{\bf x^i}})...\phi_{NL}(t,{{\bf x^N}}), \label{ansatzmany}\end{equation}where the linear wave is properly symmetrized in the case of identical particles, and contains actually all the physical features (such as e.g. entanglement) associated to the usual, linear quantum physics, which has been succesfully confirmed in a multitude of experiments in the last century. The non-linear part is not entangled however and does not respect in general symmetrisation (anti-symmetrisation) properties required in presence of bosons (fermions). Even bosonic symmetry is systematically broken because particles are supposedly located in tiny regions of space separated by distances quite larger than their extent.
We also assume that the wave function obeys a generalization of equation (\ref{nonfreeNL}):
\begin{eqnarray}
{i}\hbar\frac{\partial \Psi(t,{{\bf x^1}},{\bf x^2},...,{\bf x^i},...,{\bf x^N})}{\partial t}=-\sum_{i=1...N}\hbar^2\frac{\Delta\Psi(t,{{\bf x^1}},{\bf x^2},...,{\bf x^i},...,{\bf x^N}))}{2m_i}\nonumber\\
+V^L(t,{{\bf x}})\Psi(t,{{\bf x^1}},{\bf x^2},...,{\bf x^i},...,{\bf x^N}))+V^{NL}(\Psi)\Psi(t,{{\bf x^1}},{\bf x^2},...,{\bf x^i},...,{\bf x^N})),\label{nonfreeNLmany}
\end{eqnarray}with 

 \begin{equation}V^{NL}(\Psi)=\sum_{i}{\hbar^2\over 2m_i}{\Delta|\Psi(t,{{\bf x^i}})|\over |\Psi(t,{{\bf x}})|}-{\hbar^2\over 2m_i}{\Delta|\Psi_L(t,{{\bf x^i}})|\over |\Psi_L(t,{{\bf x}})|},\end{equation}
where we sum over elementary particles, symbolically differentiated here via the label $i$. 
 Then all the results of the previous section still apply in the present case. For instance, privilegging gausson type solutions, we find, in first approximation, a formal solution of equation (\ref{nonfreeNLmany}) of the type
\begin{equation}\Psi(t,{{\bf x^1}},{\bf x^2},...{\bf x^i}...{\bf x^N})\approx N'\cdot{R_L({{\bf x^1}},{\bf x^2},...{\bf x^i}...{\bf x^N},t)\over R_L({\bf {{\bf x_0^1}},{\bf x_0^2},...{\bf x_0^i}...{\bf x_0^N}},t)}e^{i\varphi_L({{\bf x^1}},{\bf x^2},...{\bf x^i}...{\bf x^N},t)}\cdot e^{-(\sum_{i=1...N}{A_0(i)\over 2}({\bf x^i}-{\bf x^i_0(t)})^2)}\label{formalsol}
\end{equation} with $\frac{d {\bf x^i_0}}{dt}=\frac{\hbar}{m}{\bf \bigtriangledown}_i{\varphi_L({\bf x^1}},{\bf x^2},...{\bf x^i}...{\bf x^N},t)$ and $N'=R_L({\bf {{\bf x_0^1}},{\bf x_0^2},...{\bf x_0^i}...{\bf x_0^N}},t=0)\cdot N$, making use of equation (\ref{defnorm}). The guidance equation in configuration space originally derived by de Broglie in 1926 \cite{1926} is thus fulfilled, generalizing the results of the previous section to the many particles case.

At this level, the solution (\ref{formalsol}) remains a somewhat formal solution, among others because the sizes of the solitons ${1\over \sqrt{A_0(i)}}$ are left undetermined, excepted that we implicitly assume that they are smaller than the typical size of variation of the linear (pilot) wave function. We expect however that these sizes have something to do with the masses of the elementary particles. For instance, if we accept that $3\hbar \omega_0(i)/2$, the energy of the soliton estimated on the basis of equation (\ref{S6gaussf}), with $A_x=A_y=A_z=A_0(i)$, is of the order of  $m_ic^2$, the rest-mass/energy of the particle, we find that the size of the soliton (which is of the order of $1/\sqrt{A_0(i)}$) is of the order of the Compton wavelength ($\hbar/(m_ic)$) of the associated particle, making use of the relations $\omega_0(i)=\sqrt{k_0(i)/m_i}$, $\hbar^2A^2_0(i)/2m_i=k_0(i)/2$, and $3\hbar \omega_0(i)/2$=$(3\hbar^2/2m)A_0(i)$. Even if it is a bit amazing to introduce the relativistic relation $E=mc^2$ in the present discussion, we believe that, in the case where our model makes sense, the self-focusing and self-accelerating mechanisms invoked here ought to find a justification, ultimately, in the framework of relativistic quantum field theory. Although this is out of the scope of the present paper, it would be very interesting and challenging to try for instance to connect the non-linear potential (\ref{S3}) to fundamental non-linearities characterizing the Lagrangians of gauge theories  \cite{Higgs}. It could be the case after all that the self-interaction considered here has something to do with the renormalisation procedure the scope of which being precisely to eliminate self-interaction and to focus on interactions between different particles. This is in a sense what we did here by factorizing the wave function into a ``linear'' wave function $\Psi_L$ containing the usual physics (and interactions between different particles) and a ``non-linear'' wave  function $\phi_{NL}$ associated to self-interaction.

In this perspective, the gravitational interaction, which is not renormalisable, as is well-known, could play a different role, compared to gauge interactions, as we will discuss now...
 \section{The role of gravitation: experimental proposal.}
 \subsection{The role of gravitation}
 If we accept the possible existence of a non-linearity of the type considered in the present paper, we are forced to consider seriously the logical possibility of the picture according to which particles are identified with solitons of extremely small size following the lines of flow associated to the pilot wave, in accordance with the guidance equation. In virtue of the H-theorem established by Valentini and coworkers in the framework of the pilot wave interpretation \cite{norsen2018,valentini-phd,valentini042,cost10,toruva,colin2012,abcova,efthymiopoulos3,efthymiopoulos,efthymiopoulos2,grecannales}, the distribution of positions of these solitons converges in time to the Born distribution in $|\Psi_L|^2$. This important technical result holds that we interpret in the present context $|\Psi|^2$ as a probability distribution (which is the standard, Copenhagen interpretation) or not. This is so because we are free to normalise $\Psi$ at our convenience, which will affect neither the value of its barycentre nor the value of the non-linear potential which does not depend on rescalings of $\Psi$, $\Psi_L$ or $\phi_{NL}$ by construction. The quantum H-theorem  \cite{norsen2018} implies, at this level, that our model provides an ad hoc reformulation of the standard interpretation provided we treat the position as a beable (or element of reality). Now, it is worth noting that, in accordance with the spirit of the pilot wave interpretation, there is in our model no influence at all of the soliton on the pilot wave (no feedback), which implies that a priori there is no way to put into evidence the reality of the solitons. All this is not new, it was already postulated in the de Broglie-Bohm interpretation that the particle is passively guided by the pilot wave according to the guidance condition. From this point of view the de Broglie-Bohm ontology, and the double solution program as well, merely deliver an ad hoc reformulation of the standard theory.
 
 It is our hope however that it is possible to discriminate the standard interpretation from the double solution approach in the presence of gravitational fields. 
 Nothing forbids indeed to assume that the source of the gravitational potential is the density of stuff $|\Psi|^2$. A similar intuition is actually at the origin of the Schr\"odinger-Newton equation \cite{Moeller,Rosenfeld,diosi84,penrose,penrose2,CDW,hatifi}). This means that to the difference of say electro-magnetic interactions for which standard, linear predictions have been tested with very high accuracy in the framework of QED, we shall assume here that gravity provides a feedback from the non-linear to the linear sector. From that point of view the gravitation would be ``other''. In particular it would differ from other fundamental (gauge) interactions which a priori admit a satisfactory description in the framework of linear quantum mechanics. 
 \subsection{Experimental proposal}
  The presence of a feedback from the ``particle'' (soliton) on the pilot wave makes it possible to discriminate between, at one side, the standard interpretation (where the description of the quantum system is encapsulated in the linear wave function $\Psi_L$ solely) and, at the other side, the approach followed here. To show this, let us consider a recent proposal aimed at testing the nature of the gravitational interaction at the level of elementary quantum systems.  Roughly summarized, the idea \cite{bose,vedral} is to prepare two mesoscopic systems $A$ and $B$ (for instance, diamond nanospheres (beads) with a spin 1/2 NV-center inside) in a pure factorisable spin state $|\Psi(t=0)>=(\alpha_A|+_A>+\beta_A|-_A>)(\alpha_B|+_B>+\beta_B|-_B>)$ and to let each of them fall along a humpty-dumpty \cite{humpty} Stern-Gerlach device, in such a way that during a long time $\tau$ the two spheres remain parallel to each other. Each of the 4 spin states will thus accumulate a phase, due to the gravitational interaction between the two systems, so that, after recombining the wave packets at the end of the (double) Stern-Gerlach device, the state $|\Psi(\tau)>$ of the full system is equal to 
  
  $\alpha_A\alpha_Be^{i\theta_{++}}|+_A+_B>+\alpha_A\beta_Be^{i\theta_{+-}}|+_A-_B>$+$\beta_A\alpha_Be^{i\theta_{-+}}|-_A+_B>+\alpha_A\beta_Be^{i\theta_{--}}|-_A-_B>$. Performing tomography of this state delivers information about $\theta$ and thus about the nature of the gravitational interaction between the two systems. Let us denote $d_{i,j}$ (with $i,j\in \{+,-\}$) the distance between the vertical axes along which the $i$ spin component of the $A$ system and the $j$ component of the $B$ systems parallely move. The standard, linear approach, predicts that \begin{equation}\theta^{standard}_{i,j}=\tau {Gm_Am_B\over \hbar}{1\over d_{i,j}}.\label{standard}\end{equation}

 In order to modelize the self-interaction of the soliton, let us add to the  non-linearity (\ref{S3}) a non-linear coupling {\it \`a la} Schr\"odinger-Newton  \cite{CDW}:  the self-gravitational potential is thus equal to $V^\mathrm{self-grav.}({\bf x}^{i})=-\int {d}^3{\bf x'}^{i}|\Psi(t,{\bf x'}^{i})|^2{Gm^2\over |{\bf x}^{i}-{\bf x'}^{i}|}$, where ${\bf x}^{i}$ represents the position of the $i$-particle (here the L2-norm of $\Psi(t,{\bf x}^{i})$ is normalised to 1). If the shape of the soliton is a Heaviside isotropic function of radius $R$, we find by a straightforward computation that $V^\mathrm{self-grav.}(|{\bf x}^{i}|)= \frac{Gm^2}{R}~\!\left(-\frac{3}{2}+\frac{1}{2}\left(\frac{d}{R_i}\right)^2\right) \text{if }d\leq R_i$ where $d$ denotes the distance to the center of the soliton: $d=|{\bf x}^{i}|-|{\bf x_0}^{i}|$ and $R_i$ is the Compton wavelength of the $i$ particle $R_i=\hbar/(m_i\cdot c)$.
Otherwise, for larger distances ($d$ larger than the radius $R_i$), one can integrate the internal contributions using Gauss's theorem so that: $V^\mathrm{eff}(d) =- \frac{Gm^2}{d}\quad (d\geq R_i).$

This is actually is a generic behaviour. Similar predictions can be made if for instance we assume that the shape of the soliton is not a Heaviside but a gaussian function (gausson \cite{Biali}): up to a constant of the order of $ -\frac{Gm^2}{R}$ which can be considered as a correction to the rest mass energy of the particle \cite{Lucas}, the gravitational potential is an harmonic self-focusing, potential at short scales and outside from the tiny zone where the ``stuff'' is located  we find the usual Newton potential making use of Gauss's theorem. In general, the potential is anharmonic in-between but this does not really matter because the gravitational interaction can be considered as a small perturbation compared to the non-linear potential  (\ref{S3}). Generically the spring constant of the harmonic self-gravitational potential, in the vicinity of the centre, is of the order of $G\cdot \rho_i$ where $\rho_i$ is the density of stuff evaluated at ${\bf x_0}^{i}$, the center of the soliton. In the present case, $\rho_i$ is of the order of the mass of the particle divided by the cube of its Compton wavelength: $\rho_i\approx m_i/(\hbar/m_i\cdot c)^3=m_i^4c^3/\hbar^3$ and one can check that if we consider a gausson representing an electron or a nucleon, the self-gravitational spring constant is quite smaller than the spring constant associated to the self-focusing potential ${\hbar^2\Delta |\phi_{NL}|\over 2m_i  |\phi_{NL}|}$ (\ref{S6gaussf}), because their ratio is equal to $G\cdot m_i^2/\hbar\cdot c$ which is extremely small when the particle is an electron or a nucleon. We find the same ratio between the gravitational self-energy (which is of the order of $-Gm^2_i/R_i=-Gm^2_i/(\hbar/ m_i c)$) and the self-energy $(\hbar^2/(m_iR_i^2))=m_ic^2$ associated to the self-focusing potential ${\hbar^2\Delta |\phi_{NL}|\over 2m_i  |\phi_{NL}|}$.  Making use of Ehrenfest's theorem, it is easy to show that the self-gravitational potential is a non-accelerating potential \cite{debroglieDurt}, because each cartesian component of the global force $\int {d}^3{\bf x}_{i}\int {d}^3{\bf x'}_{i}|\Psi(t,{\bf x}_{i})||\Psi(t,{\bf x'}_{i})|^2Gm^2{\bf \nabla}_{{\bf x}_{i}}{1\over |{\bf x}_{i}-{\bf x}'_{i}|}$ nullifies. This is so because it is the integral of an odd function over an even domain.

If we consider its effect on the solitons, the self-gravitational interaction is thus a small and non-accelerating potential \cite{debroglieDurt} which will reinforce the self-focusing character of the non-linear potential  (\ref{S3}). As it is non accelerating, its presence does not modify our previous analysis; in particular the guidance equation (\ref{guidance}) is still satisfied and the linear Schr\"odinger equation (\ref{S1V}) remains unaltered at this level (no feedback from the non-linear to the linear sector). However, if we consider regions of the pilot wave where the soliton is not located, and separated of the center of the soliton by a distance larger than its size, Gauss's theorem applies and we expect the soliton to interact with the pilot wave via an effective Newtonian potential. It is at this level that the feedback from the non-linear to the linear sector appears. In this approach, we also predict that the mass-energy of the elementary particles of the $A$ and $B$ systems is concentrated along the axes followed by the $k,l$ components, with a probability equal to $|\Psi_{k,l}|^2$, according to the Born rule, in virtue of the aforementioned quantum H-theorem. In this case everything happens as if the wave function associated to the center of mass of the $i,j$ component was moving inside an external gravitational potential created by the mass concentrated along the ${k,l}$ trajectory. This results in the appearance of a self-gravitational potential equal\footnote{In first approximation, the self-gravitational potential can be estimated making use of 
$V^\mathrm{eff}(d)=-Gm^2({3R^2-d^2\over 2R^3})$, which expresses the gravitational potential inside a homogeneous nanosphere of radius $R$.} to $- Gm_A^2{3\over 2R_A}$ ($-Gm_B^2{3\over 2R_B}$) along the trajectory of the $k$ ($l$) spin-component, and of the usual Newtonian potential along trajectories of spin components where the mass is not concentrated ($i\not=k$ and $l\not= j$).  Note that there also appears here a gravitational potential between the spin up (down) component and the spin down (up) component of a same nanosphere, to the difference with the standard approach.  Henceforth, we predict that with a probability equal to $|\Psi_{k,l}|^2$ (for $k,l\in \{+,-\}$), the state is a pure state such that 
\begin{eqnarray} \theta^{soliton}_{i,j}=\tau {G\over \hbar}&(&\delta{k,i}{3m_A^2\over 2R_A}+(1-\delta{k,i}){m_A^2\over d_{k^A,i^A}}+\delta{l,j}{3m_B^2\over 2R_B}+(1-\delta{l,j}){m_B^2\over d_{l^B,j^B}}\nonumber\\ &+&m_A m_B({1\over d_{i^A,l^B}}+{1\over d_{k^A,j^B}}-\delta_{k,i}\delta{l,j}{1\over d_{k^A,l^B}})) \end{eqnarray} 
 which obviously differs from the standard prediction (\ref{standard}). Moreover, the resulting state is a mixture while in the standard approach we get a pure state. Tomography of the final state would thus make it possible to falsify\footnote{Actually, a single humpty dumpty device suffices to falsify our model, because the standard theory predicts that no dephasing appears between the two branches of the humpty dumpty device when only one nanosphere is present. This is not the case with our approach, which predicts a dephasing equal to $\pm \tau {Gm^2\over \hbar} ({3\over 2R}-{1\over d})$ with $d$ the distance between the spin up and spin down paths and $R$ the radius of the nanosphere. The value of the sign of the dephasing depends on where the (mass/energy of the) nanosphere is located. The localisation of the mass/energy of the nanosphere obeys in turn the Born rule, in virtue of the quantum H-theorem.} one of the two  models, if not both  \cite{hatifi}.
\section{Conclusions}
Where is mass/energy localised? This question is as old as the quantum theory. When Einstein was struggling with the theory of the photon he already faced the following problem:  (Maxwell) waves tend to spread, but particles (photons) behave as a localised indivisible whole. He even mentioned in a correspondence with H. Lorentz \cite{letter} that maybe a non-linear generalization of Maxwell's equation would be necessary in order to solve this paradox. The problems raised by wave-particle duality in quantum wave mechanics are exactly the same. Born understood that a way to solve them was the statistical, probabilistic interpretation. There exists however another tradition in theoretical physics, that can be traced back to Poincar\'e \cite{Poincar}, according to which particles are concentrations of force fields \cite{Fer,fargue,visser}. The present paper fits to this realistic approach. It could be that, just like special relativity meant the end of aether's theories,  the quantum theory means the end of realism and that all realistic interpretations are condemned to disappear soon or late. Nevertheless they have the merit to push us to explore new mechanisms and to question the dominating orthodoxy \cite{anas}. From this point of view they let advance science. New experiments are indeed not very exciting if we know in advance their results. Testing gravitation at the microscopic and mesoscopic scales is very challenging because today no fully satisfactory quantum theoretical description of the gravitational interaction is available. The scope of the present paper is to suggest that, in this frontier domain, direct observations can provide an answer to these old questions.


\end{document}